\documentclass[a4paper]{jpconf}
\usepackage{graphicx}
\usepackage{pb-diagram,lamsarrow,pb-lams,amsfonts,amssymb,amsthm}
\usepackage{amsmath,epsfig}
\usepackage{cite}
\usepackage{slashed}
\def\be{\begin{equation}}
\def\ee{\end{equation}}
\def\bea{\begin{aligned}}
\def\eea{\end{aligned}}
\def\ba{\begin{eqnarray}}
\def\ea{\end{eqnarray}}

\def\yzero{\smash{\hbox{$y\kern-4pt\raise1pt\hbox{${}^\circ$}$}}}

\def\-{\hphantom{-}}

\def\s2{\frac{1}{\sqrt2}}

\def\IF{\relax{\rm I\kern-.18em F}}
\def\II{\relax{\rm I\kern-.18em I}}
\def\IP{\relax{\rm I\kern-.18em P}}
\def\IC{\relax\hbox{\kern.25em$\inbar\kern-.3em{\rm C}$}}
\def\IR{\relax{\rm I\kern-.18em R}}

\def\Dsl{\,\raise.15ex\hbox{/}\mkern-13.5mu D} 
\def\IZ{Z\kern-.4em  Z}

\begin{document}
\title{Mass operator of the M2-brane on a background with constant three-form}

\author{M.P. Garc\'ia del Moral$^{1,a}$, C. Las Heras$^{1,b}$, P. Le\'on $^{1,c}$, J.M. Pe\~na$^{1,d}$, A. Restuccia$^{1,e}$}

\address{$^1$Departamento de F\'{\i}sica, Universidad de Antofagasta, Aptdo 02800, Chile.}

\ead{$^a$maria.garciadelmoral@uantof.cl;
$^b$camilo.lasheras@ua.cl; $^c$pablo.leon@ua.cl; $^d$joselen@yahoo.com; $^e$alvaro.restuccia@uantof.cl}

\begin{abstract} The formulation of supermembrane theory on nontrivial backgrounds is discussed. In particular, we obtain the Hamiltonian of the supermembrane on a background with constant bosonic three form on a target space $M_9 \times T_2$.
\end{abstract}

\section{Introduction}

The supermembrane action was obtained in \cite{Bergshoeff,Bergshoeff3} using the superspace formalism, it is a generalization to 11 dimensions of the Green-Schwarz action for strings. The supermembrane, or M2-brane, is an extended bidimensional object evolving in a target space and acts as a source of supergravity in 11 dimensions. Moreover,  the kinetic terms of the five the consistent superstrings theories in 10 dimensions can be obtained from the action of the M2-brane. The gauge potential that couples to this 2-brane is the 3-form of supegravity. 

In \cite{Bergshoeff2}, the authors studied the particular case when the background is a flat superspace. They found the expressions for the supervielbeins in the worldvolume of the supermembrane and the different components of the super 3-form in this specific background. The bosonic component was set to zero. Under these assumptions, in \cite{dwhn} they were able to obtain the action and its symmetries, the equations of motion, the supercurrent, and therefore the superalgebra of the theory. It is worth to mention that the requirements that \cite{Bergshoeff2} assume, in order to obtain the components of the super 3 forms related to the flat superspace, can also be satisfied when the bosonic component is not zero but constant. This case may be relevant if we aim to study the theory with a non trivial topology in the background, which can be the case of a spacetime with compactified dimensions. 

Later on, in \cite{deWit} it was studied the hamiltonian formulation of the supermembrane evolving in a general background. They first found the bosonic part of Hamiltonian and then, with a method known as gauge completion, they  added the fermionic contributions until second order in the fermionic variables. In that work, they found that the Hamiltonian has a non trivial dependence on the non physical variable $X^-$ in the light-cone coordinates. Furthermore, any intent to solve the constraints of the theory and find $X^-$ leads to non local expression. For this reason they could not decouple, in a local way, this non physical degree of freedom from the hamiltonian. 

In this work, we study the Hamiltonian formulation of the supermembrane evolving in a Minkowski background $M_9 \times T_2$ with a constant bosonic three form. This work is organized as follows: in Section 2 we write the supermembrane action in general curved background and explain some their symmetries. Then, in Section 3 we first find the Hamiltonian formulation of the supermembrane in the particular case in which the metric of the superspace is Minkowski with a constant bosonic 3-form. We solve the problem associated with the dependence of the $X^-$ variable in the Hamiltonian and finally we write the mass operator of the theory. We also verify the consistence of the chosen background using the equation of motion of the supergravity in eleven dimensions. Finally in section 4 we present a discussion of the results.

\section{Supermembrane theory on a curved background} In this section we review well-known results found in \cite{Bergshoeff,Bergshoeff3}.
The action for a supermembrane evolving in a general background is given in terms of superspace coordinates $X^\mu(\xi)$ and $\theta^\alpha (\xi)$ 
\begin{equation}
\label{supermembrane0}
S[Z(\xi)]=\int d^3\xi \left[-\sqrt{-g(Z(\xi))}-\frac{1}{6}\epsilon^{ijk}\Pi_i^A \Pi_j^B \Pi_k^C C_{CBA} \right] \, ,
\end{equation}
where $\xi^i$ represents the world-volume coordinates, $g_{ij}=\Pi_i^{a}\Pi_{j}^{b} \eta_{{a}{b}}$ is the induced metric of the world volume ($\eta_{ab}$ is the metric of the tangent superspace) and $\Pi_i^A=\partial_iZ^M E_M^A$. Here $M=(\mu, \alpha)$ and $A=(a,\hat{\alpha})$ represents the curved superspace index and the tangent superspace index, respectively.

This action is invariant under two local gauge invariances

\begin{itemize}
\item{World-volume reparametrizations
\begin{equation}\label{2}
\delta Z^M =\eta^i(\xi)\partial_i Z^M \, .
\end{equation}
}
\item{Local fermionic transformation ($\kappa$-symmetry)
\begin{equation}\label{3}
\delta Z^M E_M^a =0, \quad \delta Z^M E^{\hat{\alpha}}_M = (1-\Gamma)^{\hat{\alpha}}_{\hat{\beta}}\kappa^{\hat{\beta}} \, ,
\end{equation}
where $\kappa(\xi)$ is an arbitrary non constant spinor and $\Gamma$ is defined by 
\begin{eqnarray}
\Gamma &=& \frac{\epsilon^{ijk}}{6\sqrt{-g}}E_i^\mu E_j^\nu E_k^\rho\Gamma_{\mu\nu\rho}\, ,
\end{eqnarray}
and it does satisfy two main identities
\begin{eqnarray}
\Gamma^2 = 1, \quad \Gamma\slashed{E_i}=\slashed{E_i}\Gamma = g_{ij}\frac{\epsilon^{jkl}}{2\sqrt{-g}}E^\mu_kE^\nu_l\Gamma_{\mu\nu}\, .
\end{eqnarray}
}
\end{itemize}
The existence of this $\kappa$-symmetry in the action imposes strong conditions under the allowed backgrounds of the theory. Specifically, in a $D=11$ superspace, the constraints imposed by the $\kappa$-symmetry over the four form field strength and the torsion two form
\begin{equation}
H=dC, \quad T^A=\frac{1}{2}E^BE^CT_{CB}^A\, ,
\end{equation}
are given by 
\begin{equation}
H_{\hat{\alpha}\hat{\beta}\hat{\gamma}\hat{\delta}}=H_{\hat{\alpha}\hat{\beta}\hat{\gamma}d}=0 \, , \label{hc1}
\end{equation}
\begin{equation}
T^a_{\hat{\alpha}\hat{\beta}}=(\Gamma^a)_{\hat{\alpha}\hat{\beta}}\, , \quad H_{\hat{\alpha}\hat{\beta}ab}=-2(\Gamma_{ab})_{\hat{\alpha}\hat{\beta}}\, , \label{hc2}
\end{equation}
\begin{equation}
\eta_{c(a}T^c_{b)\hat{\alpha}}=\eta_{ab}\Lambda_{\hat{\alpha}}\, , \quad H_{\hat{\alpha}abc}=-\frac{1}{2}\Lambda_{\hat{\beta}}(\Gamma_{abc})^{\hat{\beta}}_{\hat{\alpha}}\, . \label{hc3}
\end{equation}

Here, the $\Lambda_{\hat{\alpha}}$ is an arbitrary spinor which in eleven dimension we can be always set to zero. This constraints are related with the coupling of the supermembrane theory to the eleven dimensional supergravity.
\section{Supermembrane theory on a generalized flat superspace}
In this section we will study the Hamiltonian formulation of the supermembrane on a Minkowski space $M_9 \times T_2$  with a constant bosonic three-form and $T_2$ a flat 2-torus. This will be done using the results of the previous section and the same procedure used in \cite{dwhn}. Then considering that the metric of the target-space is given by $G_{\mu \nu}= \eta_{\mu \nu}$ the components of the supervielbein take the following form
\begin{equation}
E^a_M = (\delta^a_\mu,-(\bar{\theta}\Gamma^a)_\alpha), \quad E^{\hat{\alpha}}_M=(0, \delta^{\hat{\alpha}}_\alpha). \label{sv}
\end{equation} 
Using this expresions for the supervielbein and the constraints (\ref{hc1})-(\ref{hc3}) we can solve $dC=H$ for $C$ and obtain that the most general solution for this case is given by 
\begin{eqnarray}
C_{\mu \nu \alpha}&=&(\bar{\theta}\Gamma_{\mu \nu})_{\alpha} \, , \label{3f1}\\
C_{\mu \alpha \beta}&=&(\bar{\theta}\Gamma_{\mu \nu})_{(\alpha}(\bar{\theta}\Gamma^\nu)_{\beta )} \, , \label{3f2}\\
C_{\alpha \beta \gamma}&=&(\bar{\theta}\Gamma_{\mu \nu})_{(\alpha}(\bar{\theta}\Gamma^\mu)_{\beta}(\bar{\theta}\Gamma^\nu)_{\gamma)}\, , \label{3f3} \\
C_{\mu \nu \rho}&=& const\, . \label{3f4} 
\end{eqnarray}
Now, introducing (\ref{sv}) and (\ref{3f1})-(\ref{3f4}) in the action (\ref{supermembrane0}) it can be shown that the most general action for this particular case is given by
\begin{eqnarray}
\label{cca}
S & = & \int d^3 \xi \left\lbrace -\sqrt{-g}-\varepsilon^{ijk}\bar{\theta}\Gamma_{\mu \nu}\partial_k \theta \left[ \frac{1}{2}\partial_i X^\mu (\partial_j X^\nu +\bar{\theta}\Gamma^\nu \partial_j\theta) \right. + \right. \nonumber \\
& + & \left. \left.  \frac{1}{6}\bar{\theta}\Gamma^\mu \partial_i \theta \bar{\theta}\Gamma^\nu \partial_j \theta \right] -\frac{1}{6}\varepsilon^{ijk}\partial_i X^\mu \partial_j X^{\nu} \partial_k X^\rho C_{\rho \nu \mu} \right\rbrace. \label{cca}
\end{eqnarray}
The latter term in (\ref{cca}) is zero when the space is purely a contractible space, like Minkowski spacetime. However it has a nontrivial contribution when there is a compact sector of the target space.

As this action is a particular case of (\ref{supermembrane0}), it is invariant under reparametrizations of the worldvolume (\ref{2}) and $\kappa$-symmetry (\ref{3}), but also under super-Poincare transformations given by
\begin{equation}
\delta X^\mu = -\bar{\epsilon}\Gamma^\mu\theta +l^\mu_\nu X^\nu+a^\mu, \qquad
\delta\theta = \epsilon +\frac{1}{4}l_{\mu\nu}\Gamma^{\mu\nu}\theta\, .
\end{equation}
In order to proceed with the physical analysis of the supermembrane we will pass to the Light-Cone coordinates.
We will denote the transverse coordinates by $X^a$ with $a=1,..,9$. Then we can make a partial use of the worldvolume reparametrizations and the $\kappa$-symmetry to perform the following gauge fixings 
\begin{equation}
X^+(\xi)= X_0^+ +\tau, \quad \Gamma^+\theta = 0. 
\end{equation}
Using now this expressions and following the same procedure applied in \cite{dwhn} we find that the Hamiltonian density is given by

\begin{equation}
\mathcal{H}=\frac{1}{P_--C_-}\left[\frac{1}{2}(P_a-C_a)^2 +\frac{1}{4}(\varepsilon^{rs}\partial_rX^a\partial_sX^b)^2\right] +\varepsilon^{rs}\bar{\theta}\Gamma^-\Gamma_a\partial_s\theta \partial_r X^a-C_+ -C_{+-}\, 
\end{equation}
where
\begin{eqnarray}
C_a &= & -\varepsilon^{rs}\partial_r X^- \partial_s X^b C_{-ab}+\frac{1}{2}\varepsilon^{rs}\partial_r X^b \partial_s X^c C_{abc} \, ,\\
 C_{\pm} &= & \frac{1}{2}\varepsilon^{rs}\partial_r X^a \partial_sX^bC_{\pm ab} \, ,\\
 C_{+-} &= & \varepsilon^{rs}\partial_rX^-\partial_sX^aC_{+-a}\, .
\end{eqnarray}
This hamiltonian density is subject to the following primary constrains
\begin{eqnarray}
\Phi_r & \equiv & P_a\partial_r X^a+P_-\partial_r X^- +\bar{S}\partial_r \theta \approx 0\, , \label{ch1} \\
\chi & \equiv & S+ (P_--C_-) \Gamma^-\theta \approx 0\, .
\end{eqnarray}
It is an easy calculation to verify that there are no secondary constraints in the theory. 

Now, we can use the tensor gauge transformation of the three from to fix the gauge $C_{+-}=0 \,, C_{-ab}=constant$.  Nevertheless this Hamiltonian density has an explicit and non trivial dependence on the $X^-$ variable, which is not desirable. In fact if we try to solve for $X^-$ in (\ref{ch1}) this will lead us to non local expressions. This problem was reported for the first time in \cite{deWit}.

In order to solve this problem, we propose the following transformation of the canonical variables
\begin{equation}
\begin{array}{ll}
X^a \rightarrow \hat{X}^a \equiv X^a, & X^- \rightarrow \hat{X}^- \equiv X^- \,,\\
P_a \rightarrow \hat{P}_a \equiv  P_a - C_a \,,& P_- \rightarrow \hat{P}_- \equiv  P_- -C_- \,,\\
\theta^\alpha \rightarrow \hat{\theta}^\alpha \equiv \theta^\alpha\,, &  S \rightarrow \hat{S} \equiv  S\,.
\end{array} \label{ct}
\end{equation}
It can be shown, that this transformation does preserve all the Poisson brackets and also satisfy 
\begin{equation}
\int_{\Sigma} (P_a \dot{X}^a +P_{-}\dot{X}^{-}+ \bar{S} \dot{\theta} ) = \int_{\Sigma} ({\hat{P}}_a \dot{\hat{X}}^a +{\hat{P}}_{-}\dot{\hat{X}}^{-}+ \hat{\bar{S}} \dot{\hat{\theta}})\,.
\end{equation}
Then we can conclude that this is a canonical transformation. 

In this new canonical variables the Hamiltonian density is given by
\begin{equation}
\mathcal{\hat{H}}=\frac{1}{\hat{P}_-}\left[\frac{1}{2}\hat{P}_a\hat{P}^a +\frac{1}{4}(\varepsilon^{rs}\partial_r\hat{X}^a\partial_s\hat{X}^b)^2\right] +\varepsilon^{rs}\bar{\hat{\theta}}\Gamma^-\Gamma_a\partial_s\hat{\theta} \partial_r \hat{X}^a-\hat{C}_+\,,
\end{equation} 
subject to the constraints 
\begin{eqnarray}
\hat{\Phi}_r & \equiv & \hat{P}_a\partial_r \hat{X}^a+\hat{P}_-\partial_r \hat{X}^- +\hat{\bar{S}}\partial_r \hat{\theta} \approx 0\,, \label{ch1} \\
\hat{\chi} & \equiv & \hat{S}+ \hat{P}_- \Gamma^-\hat{\theta} \approx 0\,.
\end{eqnarray}
Although the fermionic constraint change it structure after we perform the canonical transformation, it is easy to see that the Poisson Bracket with itself remain invariant. 

Finally we can also fix the gauge 
\begin{equation}
P_-=\sqrt{w(\sigma)}, \quad \mbox{with} \int_{\Sigma}\sqrt{w(\sigma)}=1,
\end{equation}
which, together with the definitions of the zero modes 
\begin{eqnarray}
\hat{P}_0^a  =  \int_\Sigma d^2\sigma \hat{P}^a, \quad \hat{P}_0^-=- \int_{\Sigma} d^2\sigma \mathcal{\hat{H}},   
\end{eqnarray}
\begin{equation}
\hat{X}_0^a =  \int_\Sigma d^2 \sigma \sqrt{w(\sigma)}\hat{X}^a, \quad \hat{\theta_0} =  \int_\Sigma d^2 \sigma \sqrt{w(\sigma)} \hat{\theta},
\end{equation}
allow to us find the mass operator of the supermembrane in this background,
\begin{eqnarray}
\mathcal{M}^2 \equiv -2P_0^+P_0^--\hat{P}_{0a}\hat{P}^a_0 & = & \int_\Sigma d^2\sigma \left\lbrace \frac{1}{\sqrt{w(\sigma)}}\left[\hat{P}'_a \hat{P}'^a + \frac{1}{2}(\varepsilon^{rs}\partial_r\hat{X}'^a\partial_s\hat{X}'^b)^2 \right] \right.+ \nonumber \\ & + & \left. 2\varepsilon^{rs}\hat{\bar{\theta}}'\Gamma^-\Gamma_a\partial_s\hat{\theta}' \partial_r \hat{X}'^a -2\hat{C}_+ \right\rbrace
\end{eqnarray}
where the prime indicates that we are excluding the zero modes contribution and we are considering that $\theta^{'}$ is single-valued, even in a compactified target-space. The fact that $\hat{\theta}_0$ does not appear in the mass operator has a very important consequence for the theory, that where discussed in detail in \cite{dwhn}. In that  work the authors used zero modes independence in the mass operator to prove that, if there exist a massless states for the supermembrane, it would be the massless supermultiplet of the eleven dimensional supergravity. 

\subsection{Consistency of the background}
It may be questioned if $G_{\mu\nu}=\eta_{\mu\nu}$ with $C_{\mu\nu\rho}=constant\neq 0$ is a background consistent with supergravity. This background was already considered in  \cite{Duff4}. Let us consider the bosonic action of supergravity in $d=11$ \cite{Cremmer}, the equation of motion for the elfbein is \cite{Cremmer2}
\begin{eqnarray}
R_{\mu\nu} - \frac{1}{2}g_{\mu\nu}R = \frac{1}{3}g_{\mu\nu}F_{\rho\sigma\lambda\tau}F^{\rho\sigma\lambda\tau} - \frac{1}{24}F_{\mu\rho\sigma\lambda}F_\nu^{\rho\sigma\lambda}\,,
\end{eqnarray}
and, is easy to see that the scalar curvature is given by
$
R=\frac{1}{36}F_{\rho\sigma\lambda\tau}F^{\rho\sigma\lambda\tau}\,.
$
As the bosonic components of the super 3 form are set to constant, therefore the field strength $F=dC$ is zero. We have then, that the equation of motion can be written as
$
R_{\mu\nu}=0\,,
$
and we have that the vacuum solution must be Minkowski in 11 dimensions. It can be verified that this is not only a maximally symmetric space, but a maximally supersymmetric space \cite{Figueroa}, because of the vanishing condition for the field strength.

\section{Discussion}
We present in this work the action for a supermembrane evolving in a target space with constant bosonic 3-form. The consequences of this study  acquire relevance in the case where there are compactified dimensions in the target space, we considered in our analysis $M_9\times T_2$. This study enable us to obtain the Hamiltonian  of the supermembrane in this particular background. We found the same problem reported in \cite{deWit} with respect to the nontrivial dependence of the Hamiltonian on the variable $X^-$. We solved this problem by performing a canonical transformation, which eliminates the $X^-$ dependence in the Hamiltonian by solving the first class constraint of the theory. We found in the usual way $X^-$ in terms of the physical degrees of freedom. Once the Hamitonian is expressed in terms of these  variables, we compute the Mass operator of the theory and we find that it does not depend of the zero modes of the theory.   

Using this new well defined Hamiltonian it is possible to study in further detail the properties of the supermembrane in a constant curved background, like for example, the asymptotic coupling to the  solution of supergravity in eleven dimensions obtained in \cite{Duff4} with a M2-brane acting as a source.  

\section{Acknowledgements}  A.R. and M.P.G.M. are partially supported by the Project Fondecyt 1161192 (Chile). J.M.P, C.L.H y P.L are  supported by the Project MINEDUC-UA codes ANT1855 and ANT1856 of the Universidad de Antofagasta.

%
\section*{References}

\providecommand{\newblock}{}

\end{document}